\documentclass{vmax-preprint}

\usepackage{listings}
\usepackage{tikz}
\usetikzlibrary{positioning,arrows,shapes.geometric,fit,calc}
\graphicspath{{figures/}{figs/}}
\newcommand{\BASEPCT}{11.6\%}

\newcommand{\GRPOBASEPCT}{27.6\%}
\newcommand{\GRPOSFTPCT}{43.6\%}
\newcommand{\GRPOSFTDELTA}{$+16.0$}

\title{unix-ctf: Procedural Environments for Unix-Competence Reinforcement Learning}
\subtitle{A procedural training surface for shell and operating-system competence}

\author{Geoffrey Bradway}
\author{Roger Creus Castanyer}
\author{Lorenz Wolf}
\author{Maxwill Lin}
\author{Matthew James Sargent}
\author{Augustine N. Mavor-Parker}
\affiliation{Vmax}

\abstract{\textit{Unix competence} is the ability to use shell and operating-system
primitives as first-class tools, not merely to write programs through a
terminal. Current terminal benchmarks tend to blur this distinction: a solver
fluent in Python but weak in Unix can pass a substantial fraction of
Terminal-Bench~2.0, while the reverse skill profile is rarely
exercised.
We make the distinction operational and build a training surface for the
Unix component. \textbf{unix-ctf} is a procedural generator of
capture-the-flag tasks for shell agents. Each task hides a short token
(a \emph{flag} of the form \texttt{flag\{a3b1c9...\}}) inside a fresh
Linux container using a single Unix feature, and the agent must recover
it. Tasks are produced by an LLM-assisted synthesis pipeline that
generates candidate hiding techniques, rewrites them into parameterized
hide-and-find script pairs, and filters them with a bidirectional
contract: the hide script must leave no plaintext trace of the flag on
disk, and the find script must recover the flag in a fresh directory.
Because the LLM only writes the planting and recovery steps (the
container, layout, and grading harness are fixed), the pipeline lands
656 of 750 raw attempts as portable, reusable variants
(\textbf{87.5\%}). Our reproduction of Endless Terminals'
full-container-generation approach lands only 17.4\% under the same
checks. The 656 variants canonicalize to 155 distinct techniques. Fine-tuning Qwen3-8B with LoRA using GRPO on this surface
lifts solve rate from \BASEPCT{} to \textbf{\GRPOSFTPCT{}} on a
15-skill multi-family holdout ($n{=}225$), redistributes which
InterCode-CTF tasks the model solves, and produces a $+33$\,pp gain in
Forensics while reaching $32/100$ on InterCode-CTF. These results
suggest that Unix competence is separable, trainable, and best evaluated
directly rather than folded into programming-through-a-shell.
}

\date{\today}
\correspondence{\emails{gb, matthew, augustine}}

\begin{document}
\maketitle

\section{Introduction}
\label{sec:intro}

Terminal interaction has become a standard setting for evaluating
language-model agents. InterCode~\citep{yang2023intercode},
Terminal-Bench~\citep{merrill2026terminalbench},
AgentBench-OS~\citep{liu2024agentbench}, NYU CTF
Bench~\citep{shao2024nyuctf}, and Cybench~\citep{zhang2024cybench}
are now common evaluation targets, and large-scale training pipelines
such as Endless Terminals~\citep{gandhi2026endless} and
Nemotron-Terminal~\citep{pi2026nemotron} have followed. This landscape
is usually described as measuring a single capability: using a shell to
accomplish a task on a computer. We argue that this framing collapses two different skills
into one. One is general programming performed through a terminal; the
other is Unix competence itself. This paper isolates the latter and
builds a generator for the part of the surface that current pipelines
mostly miss.

We use the \emph{capture-the-flag} (CTF) format inherited from security
competitions~\citep{shao2024nyuctf,zhang2024cybench}. Each task hides a
short token, or \emph{flag}, of the form
\texttt{flag\{a3b1c9...\}} inside a fresh sandboxed environment; the
agent must find it and print it back. The format is useful because the
reward is mechanical and unambiguous, and because a planted flag can be
tied to a single Unix feature. Recovering the flag then requires using
that feature rather than solving a loosely specified programming task.

\subsection{Unix-competence versus coding-in-a-terminal}

We define a task as testing \textbf{Unix competence} when its solution
depends on an OS, shell, or file-format feature with no clean analogue
in ordinary general-purpose programming. Concretely: a flag stored as
the value of a POSIX extended attribute (\texttt{user.secret}) on a
decoy file is invisible to \texttt{ls\ -l}, \texttt{stat}, \texttt{file},
or a recursive \texttt{grep} of the file's bytes, because \texttt{xattrs} live in
a side-channel inode namespace; the canonical recovery is a single
\texttt{getfattr -n user.secret decoy.txt}. Using a programming language
like Python does not remove the dependency: \texttt{os.getxattr} is just
a wrapper over the \texttt{getxattr(2)} syscall. The OS-specific access
path is irreducible. Other library instances (the GNU build ID inside an
ELF \texttt{.note.gnu.build-id} section, files identified by pre-epoch
\texttt{mtime}, custom OIDs inside an X.509 certificate) have the
same shape; six worked exemplars are in Appendix~\ref{app:exemplars}.

We define a task as \textbf{coding-in-a-terminal} when the deliverable
is a non-trivial program, such as a small neural network, an interpreter, a
distributed-training primitive, or a statistical sampler, and the shell
is only the medium through which the program is created, compiled, and
tested. In this regime the shell contributes little to success: a solver
fluent in Python but weak in Unix can still perform well.

Both skills are legitimate, but they lie on substantially different
axes. Treating them as one obscures what a benchmark measures and what a
training pipeline teaches. We quantify the resulting bias in current
terminal benchmarks in \S\ref{sec:related} (Figure~\ref{fig:taxonomy})
and target the underweighted Unix-competence half.

\begin{contributions}
\item \textbf{A pipeline factoring that yields procedural Unix tasks at
$5\times$ the rate of comparable LLM-driven generators.} unix-ctf
separates a pre-debugged \texttt{ctf-base} image from per-technique
planting logic, restricting the LLM to the hide-and-find script layer
that can be validated by a bidirectional contract: no plaintext flag on
disk, and recovery in a fresh directory. This produces 656 portable
variants from 750 raw attempts (\textbf{87.5\%}) versus 17.5\% in our
Endless Terminals reproduction ($n{=}120$, where 71.7\% of failures
occur at the Docker-build or test-validation stages this factoring
removes), canonicalizing to 155 IDs. See~\S\ref{sec:library}.

\item \textbf{A procedural environment generator.} The generator
instantiates fresh Docker containers from the library, plants eight
flags per container to provide a denser reward signal than a single
binary outcome, and rotates a sticky pool of fifty techniques during
training so the agent sees a broad slice of the library across batches.
See~\S\ref{sec:pipeline}.

\item \textbf{Training as a probe of learnability.} Forty batches of
GRPO on Qwen3-8B lift solve rate
\BASEPCT{}\,$\to$\,\GRPOSFTPCT{} on a 15-skill multi-family holdout
($3.8\times$, $n{=}225$). On InterCode-CTF, training redistributes
\emph{which} tasks the model solves, with the largest category gain
(\textbf{$+33$\,pp}) in Forensics. See~\S\ref{sec:results}.
\end{contributions}

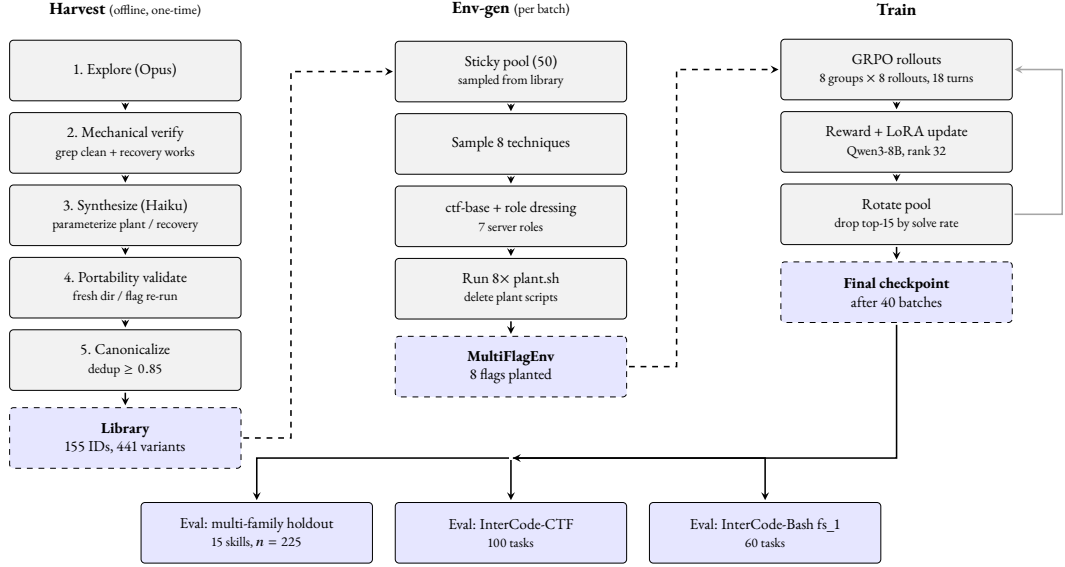
\begin{figure}[t]
\centering
\resizebox{\textwidth}{!}{%
\begin{tikzpicture}[
  font=\footnotesize,
  >=stealth,
  every node/.style={align=center},
  stage/.style={draw, rounded corners=2pt, minimum width=42mm, minimum height=11mm, fill=black!5, inner sep=2pt},
  output/.style={draw, dashed, rounded corners=2pt, minimum width=42mm, minimum height=11mm, fill=blue!10!white, inner sep=2pt},
  evalbox/.style={draw, rounded corners=2pt, minimum width=42mm, minimum height=11mm, fill=blue!10!white, inner sep=2pt},
  phase/.style={draw=none, font=\small\bfseries},
  arr/.style={->, thick, shorten >=1pt, shorten <=1pt},
  cross/.style={->, thick, dashed, shorten >=1pt, shorten <=1pt},
  loop/.style={->, thick, gray!70, shorten >=1pt, shorten <=1pt}
]

\def\xA{0mm}
\def\xB{70mm}
\def\xC{140mm}

\node[phase] at (\xA,0) (pA) {Harvest \textnormal{\scriptsize (offline, one-time)}};
\node[phase] at (\xB,0) (pB) {Env-gen \textnormal{\scriptsize (per batch)}};
\node[phase] at (\xC,0) (pC) {Train};

\node[stage, below=3mm of pA] (a1) {1. Explore (Opus)};
\node[stage, below=2mm of a1] (a2) {2. Mechanical verify\\\scriptsize grep clean + recovery works};
\node[stage, below=2mm of a2] (a3) {3. Synthesize (Haiku)\\\scriptsize parameterize plant / recovery};
\node[stage, below=2mm of a3] (a4) {4. Portability validate\\\scriptsize fresh dir / flag re-run};
\node[stage, below=2mm of a4] (a5) {5. Canonicalize\\\scriptsize dedup $\geq 0.85$};
\node[output, below=3mm of a5] (aOut) {\textbf{Library}\\155 IDs, 441 variants};

\node[stage, below=3mm of pB] (b1) {Sticky pool (50)\\\scriptsize sampled from library};
\node[stage, below=2mm of b1] (b2) {Sample 8 techniques};
\node[stage, below=2mm of b2] (b3) {ctf-base + role dressing\\\scriptsize 7 server roles};
\node[stage, below=2mm of b3] (b4) {Run 8$\times$ plant.sh\\\scriptsize delete plant scripts};
\node[output, below=3mm of b4] (bOut) {\textbf{MultiFlagEnv}\\8 flags planted};

\node[stage, below=3mm of pC] (c1) {GRPO rollouts\\\scriptsize 8 groups $\times$ 8 rollouts, 18 turns};
\node[stage, below=2mm of c1] (c2) {Reward + LoRA update\\\scriptsize Qwen3-8B, rank 32};
\node[stage, below=2mm of c2] (c3) {Rotate pool\\\scriptsize drop top-15 by solve rate};
\node[output, below=3mm of c3] (cOut) {\textbf{Final checkpoint}\\after 40 batches};

\foreach \i/\j in {a1/a2,a2/a3,a3/a4,a4/a5,a5/aOut} \draw[arr] (\i) -- (\j);
\foreach \i/\j in {b1/b2,b2/b3,b3/b4,b4/bOut} \draw[arr] (\i) -- (\j);
\foreach \i/\j in {c1/c2,c2/c3,c3/cOut} \draw[arr] (\i) -- (\j);

\draw[loop] (c3.east) -- ++(9mm,0) |- (c1.east);

\draw[cross] (aOut.east) -- ++(9mm,0) |- (b1.west);
\draw[cross] (bOut.east) -- ++(9mm,0) |- (c1.west);

\node[evalbox] at (\xB, -95mm) (eCenter) {Eval: InterCode-CTF\\\scriptsize 100 tasks};
\node[evalbox, left=4mm of eCenter] (eLeft) {Eval: multi-family holdout\\\scriptsize 15 skills, $n=225$};
\node[evalbox, right=4mm of eCenter] (eRight) {Eval: InterCode-Bash fs\_1\\\scriptsize 60 tasks};

\coordinate (junc) at ($(eCenter.north)+(0,8mm)$);
\draw[arr] (cOut.south) |- (junc);
\draw[arr] (junc) -| (eLeft.north);
\draw[arr] (junc) -- (eCenter.north);
\draw[arr] (junc) -| (eRight.north);

\end{tikzpicture}
}
\caption{The unix-ctf pipeline. \textbf{Harvest} (left, run offline once) builds a canonicalized library of 155 techniques and 441 variants through a five-stage cycle that alternates frontier-model exploration, mechanical verification, smaller-model synthesis, portability validation, and dedup-based canonicalization. \textbf{Env-gen} (centre, run every training batch) draws a sticky pool of 50 techniques from the library, samples eight per batch, and assembles a single \texttt{MultiFlagEnv} on top of a pre-built \texttt{ctf-base} container dressed as one of seven server roles. \textbf{Train} (right) runs forty batches of GRPO against the env stream, rotating the pool after each batch (grey loop arrow). The final checkpoint is then evaluated on three benchmarks (bottom row). Dashed arrows denote inter-stage data flow.}
\label{fig:pipeline}
\end{figure}

\section{Related work}
\label{sec:related}

We position unix-ctf against four adjacent lines of work. The full
unix-ctf library composition by family appears in Appendix~\ref{app:family}.

\begin{figure}[t]
  \centering
  \includegraphics[width=\linewidth,trim=0 0cm 0 0,
        clip]{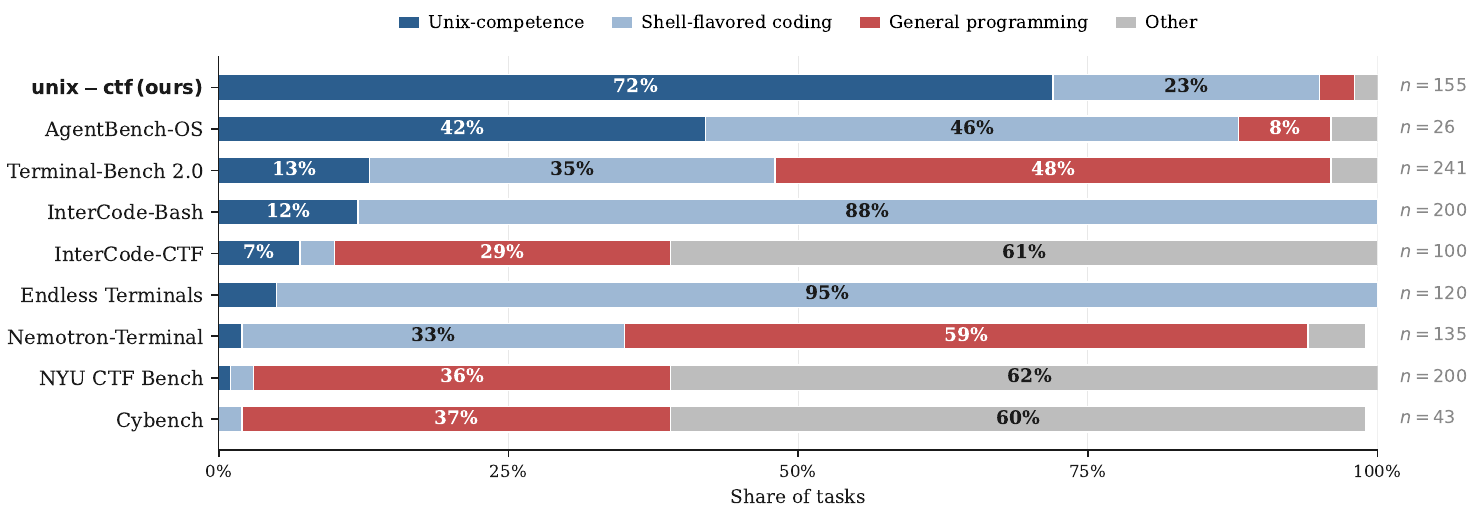}
  \caption{What terminal benchmarks actually measure. Each row: share of tasks classified by \texttt{gpt-5.4-nano} (full rubric, Appendix~\ref{app:taxonomy_rubric}). \textit{Unix-competence}: non-obvious OS-/format-internal feature with no clean Python analogue. \textit{Shell-flavored coding}: glue using \texttt{find/grep/awk/sed/sqlite3/tar} (Python solve scripts wrapping such glue count here, not as general programming). \textit{General programming}: non-trivial programs (ML, interpreters, cryptanalysis). \textit{Other}: not primarily shell/OS/programming. Endless Terminals: 120 LLM outputs from our reproduction. Nemotron: 135-task sample (15$\times$9 \texttt{skill\_based/easy} categories; harder tiers may differ). The unix-ctf row is classified by us against our own corpus; the other eight rows are external benchmarks whose tasks were passed verbatim through the same single-task LLM rubric, without per-benchmark calibration or revision.}
  \label{fig:taxonomy}
\end{figure}

\paragraph{What current terminal benchmarks measure.}
Figure~\ref{fig:taxonomy} classifies the public task lists of nine
terminal benchmarks, together with our library, using the
Unix-competence vs.\ coding-in-a-terminal definition from
\S\ref{sec:intro}. Across Terminal-Bench~2.0, Cybench, NYU CTF Bench,
AgentBench-OS, InterCode-Bash, and Nemotron-Terminal, the
Unix-competence share rarely exceeds 30\%. The remainder is general
programming or shell-flavored composition of tools such as
\texttt{find}, \texttt{grep}, \texttt{awk}, and \texttt{sed}, usually
with a short Python equivalent. SWE-bench~\citep{jimenez2024swebench}
is even more specialized: it is Python-repository patching, with the
terminal mostly serving as the way to invoke tests. The same bias
propagates into procedural-generation pipelines. Endless Terminals
seeds scenarios such as ``developer organizing files'' or ``DevOps
engineer debugging logs,'' which are Unix-flavored but shallow: an
agent trained on this material sees \texttt{grep}, \texttt{awk}, and
\texttt{find}, but much less of the filesystem metadata, binary
formats, kernel interfaces, and service-configuration surfaces that
matter in real Unix administration. Nemotron-Terminal builds on a
primitive-skills taxonomy, such as file I/O, string manipulation, and
graph traversal, that is explicitly cross-language. The most distinctive
shell capabilities are therefore underrepresented in both evaluation
and training data, and progress on terminal benchmarks can track
programming proficiency more than Unix fluency. This is the gap we
target.

The taxonomy above motivates how we position unix-ctf against four
adjacent strands of prior work, in increasing order of structural
overlap with our setting.

\paragraph{Fixed Linux benchmarks.} The most direct evaluation
yardsticks for shell and operating-system agents are InterCode,
InterCode-CTF, Terminal-Bench~2.0, AgentBench-OS, NYU CTF Bench,
Cybench, and OSWorld~\citep{xie2024osworld}. They are hand-curated and
fixed in size (Terminal-Bench~2.0 has 89 tasks, Cybench 40, NYU CTF
Bench 200), so none exposes a generative task surface that an RL
loop can draw new instances from.

\paragraph{Computer-use benchmarks beyond the terminal.} A parallel
line, including WebArena \citep{zhou2023webarena},
WorkArena~\citep{drouin2024workarena},
OSWorld~\citep{xie2024osworld},
AndroidWorld~\citep{rawles2024androidworld}, and Windows Agent
Arena~\citep{bonatti2024windowsagentarena}, also evaluates agents in
executable software environments with programmatic success checks. The
substrate is web, desktop, or mobile control rather than the Unix
shell. unix-ctf keeps the executable-environment / mechanical-reward
shape but restricts the surface to text-only Linux interaction and
Unix-specific OS features.

\paragraph{Procedural generation in the shell and CTF space.} Our
closest precedents share both the procedural-generation goal and the
shell substrate. Endless Terminals~\citep{gandhi2026endless}
constructs 3{,}255 tasks using an OpenAI~o3~\citep{openai2025o3}
solvability filter; Nemotron-Terminal's
Terminal-Task-Gen~\citep{pi2026nemotron} produces roughly 366{,}000
trajectories from a cross-language primitive-skills taxonomy; and
CTF-Dojo / CTF-Forge~\citep{zhuo2025ctfdojo} provides 658 Dockerized
challenges extracted from \texttt{pwn.college}. All three put a single
task in each container and \textit{rely on a single-shot generator} rather than a pre-validated skill library or an asymmetric-model harvest; CTF-Dojo
in particular gives flag-based reward, but only by re-packaging
existing CTFs rather than by generating new techniques. Adjacent
procedural-generation work in the Python-repository space
(R2E-Gym~\citep{jain2025r2egym}, SWE-smith~\citep{yang2025swesmith},
SWE-Gym~\citep{pan2025swegym}) is structurally related but trains
on a different skill surface.

\paragraph{Skill libraries for language-model agents.} Voyager
\citep{wang2023voyager} for Minecraft, SkillWeaver
\citep{zheng2025skillweaver} and WALT~\citep{prabhu2025walt} for the
web, and OS-Copilot / FRIDAY~\citep{wu2024oscopilot} for operating-system
agents all share the goal of accumulating reusable skills. In
OS-Copilot and FRIDAY, skills are inferred online by the agent rather
than harvested offline. To our knowledge, unix-ctf is the first
explicit, offline-harvested, mechanically validated skill library for
Linux shell use. Self-Challenging Agents~\citep{zhou2025sca} share a
Code-as-Task contract (instruction, executable evaluator, solution), but
generate tasks per episode rather than canonicalizing them into a
reusable library.\looseness=-1

\section{The technique library}
\label{sec:library}

\begin{figure}[t]
  \centering
  \includegraphics[width=\linewidth,trim=0 .0cm 0 0,
        clip]{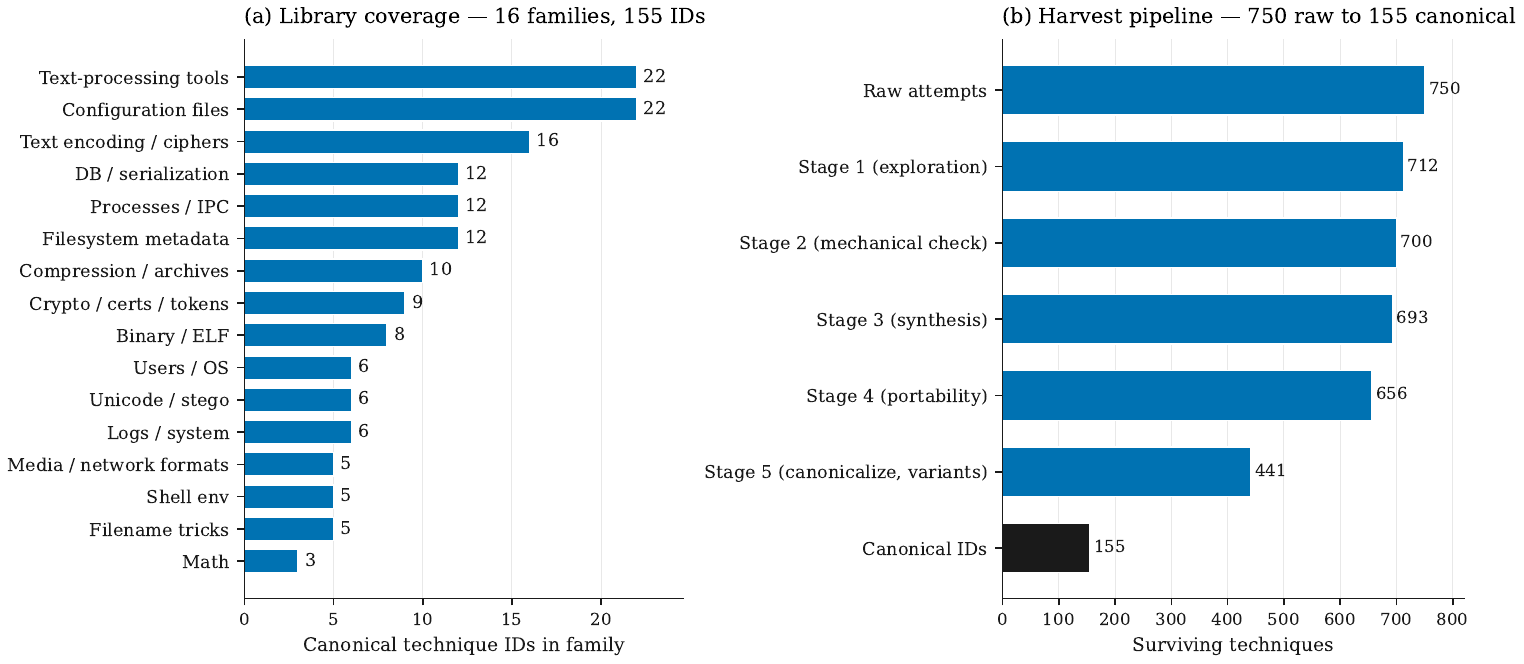}
  \caption{The technique library. \textbf{(a)} Coverage of the Unix-competence surface across 16 post-hoc families. The largest families (configuration files, text-processing tools, text encoding/ciphers) are general-purpose surfaces; the smaller families (binary/ELF, X.509, IPC primitives) are exactly the corners that distinguish Unix fluency from coding. \textbf{(b)} Production funnel from 750 raw harvest attempts to 656 portable scripts ($87.5\%$ end-to-end) and 441 deduplicated variants, canonicalized to 155 IDs. The pipeline factoring and bidirectional contract together drive the $5\times$ yield gap over Endless Terminals (\S\ref{sec:contract}).}
  \label{fig:library}
  \vspace{-1ex}
\end{figure}

Each technique enters the library through an offline pipeline that
alternates language-model generation with mechanical verification. The
result is a reusable library of \textbf{155 canonical technique
identifiers} ($k=441$ deduplicated variants), summarized in
Figure~\ref{fig:library}.

\subsection{Harvest pipeline}
\label{sec:harvest}

The five stages are designed to keep the language model on the layer it
can validate, while letting mechanical checks enforce the task contract.
(1)~\textit{Exploration:} Claude Opus 4.7~\citep{anthropic2026opus47}
runs inside a fresh \texttt{ctf-base} Docker container and is asked to
hide a flag of the form \texttt{flag\{<8--16 hex bytes>\}} using a
target technique. It emits shell commands turn by turn for up to ten
turns; we record both the command sequence and a candidate recovery
command. (2)~\textit{Mechanical verification:} in the same container, a
recursive search for the literal flag must return no occurrences on
disk, and the candidate recovery command must print the flag and exit
zero. Any failure drops the attempt. (3)~\textit{Synthesis:} Claude
Haiku 4.5~\citep{anthropic2025haiku45} rewrites the validated
exploration into two parameterized scripts, \texttt{plant.sh} (target
directory, flag) and \texttt{recovery.sh} (target directory), as a pure
textual rewrite with no shell execution. (4)~\textit{Portability
validation:} the synthesized scripts are re-run against a fresh
temporary directory with a fresh flag. \texttt{plant.sh} must exit zero,
a recursive search must find no occurrence of the new flag, and
\texttt{recovery.sh} must print the new flag on stdout and exit zero.
This catches the main synthesis failure mode: inlining the original
directory or flag. (5)~\textit{Canonicalization:} per-attempt records
are archived to \texttt{data/techniques/<id>.jsonl}, and a greedy
\texttt{difflib.SequenceMatcher} deduplication at threshold 0.85,
preferring fewer dependencies and shorter scripts, produces the
canonical \texttt{data/library/<id>.json}.

\subsection{Pipeline factoring and the bidirectional contract}
\label{sec:contract}

The 87.5\% yield is driven by two choices: \textit{what} the LLM
produces, and \textit{how} the output is filtered. Endless Terminals
asks the LLM to generate the Dockerfile, setup script, planting logic,
and tests for each task (essentially a complete container). In our
$n{=}120$ reproduction (17.5\% portable),
\textbf{$71.7\%$ of failures occur at the Docker-build or
test-validation stage}, and only $10.8\%$ at the LLM task-generation
step. The model is competent at inventing hiding techniques but
rarely succeeds at the surrounding infrastructure re-derived from
scratch each time. We therefore lift infrastructure out of the
per-attempt budget: a pre-debugged \texttt{ctf-base} image handles the
system layer, and the LLM produces only the parameterized
\texttt{plant.sh}~/~\texttt{recovery.sh} pair. On top of that we apply
two complementary filters: a recursive search of the filesystem must
\emph{not} find the flag in plaintext, and a re-run of
\texttt{recovery.sh} on a fresh directory with a fresh flag
\emph{must} recover it. Upstream pipelines use only one half: Endless
Terminals filters on whether at least one of sixteen
OpenAI~o3~\citep{openai2025o3} samples solves; SWE-smith and R2E-Gym
filter on whether a generated change invalidates a unit test. Both
check solvability without checking non-triviality.

\subsection{Quality of surviving tasks}
\label{sec:quality_audit}

Throughput alone does not establish task quality. We audit the 120
Endless Terminals task directories from our reproduction and 441
unix-ctf variants using a fixed five-dimension rubric
(Appendix~\ref{app:audit_rubric}). Three LLM judges from separate
vendors score each task independently: \texttt{gpt-5.4-nano}
\citep{openai2026gpt54nano}, \texttt{grok-4-20-reasoning}, and
\texttt{Kimi-K2.5}. We deliberately exclude Claude because it appears in
the unix-ctf generation loop.

\begin{table}[H]
\centering
\caption{Task-quality audit, mean $\pm$ std across three LLM judges from three vendors (\texttt{gpt-5.4-nano}, \texttt{grok-4-20-reasoning}, \texttt{Kimi-K2.5}; 1 worst, 5 best). 439 unix-ctf variants vs.\ 120 LLM-output task dirs from our Endless Terminals reproduction (intersection of tasks scored cleanly by all three judges). Grading-robustness dimension omitted; mechanical count: 99\% of Endless Terminals tasks have $\geq 1$ hardcoded exact-string assertion vs.\ 0\% unix-ctf. Pairwise Pearson $r$ on per-task aggregate scores (\texttt{gpt}$\times$\texttt{grok} $=0.82$, \texttt{gpt}$\times$\texttt{kimi} $=0.83$, \texttt{grok}$\times$\texttt{kimi} $=0.93$, $n{=}559$) confirms the three judges are measuring a common construct. Per-judge tables in Appendix~\ref{app:audit_rubric}.}
\label{tab:audit}
\small
\setlength{\tabcolsep}{6pt}
\begin{tabular}{lccc}
\toprule
Dimension & unix-ctf ($n{=}439$) & Endless Terminals ($n{=}120$) & $\Delta$ \\
\midrule
Input richness          & \textbf{$4.74 \pm 0.45$} & $1.81 \pm 0.32$ & $+2.93$ \\
Solution non-triviality & \textbf{$3.45 \pm 0.34$} & $2.54 \pm 0.46$ & $+0.90$ \\
Format independence     & \textbf{$4.42 \pm 1.00$} & $1.07 \pm 0.04$ & $+3.35$ \\
Tool surface diversity  & \textbf{$3.44 \pm 0.83$} & $2.85 \pm 0.78$ & $+0.59$ \\
Discoverability         & \textbf{$3.05 \pm 0.10$} & $1.12 \pm 0.08$ & $+1.93$ \\
\midrule
\textbf{Aggregate} & \textbf{$3.82 \pm 0.43$} & $1.88 \pm 0.30$ & $\mathbf{+1.94}$ \\
\bottomrule
\end{tabular}
\end{table}

\subsection{Production rates}
\label{sec:production}

On the most recent pipeline run, 712 of 750 raw harvest attempts
survived stage~1 (exploration), 700 survived stage~2 (mechanical
verification), 693 survived stage~3 (synthesis), and \textbf{656
survived stage~4 (portability)}, giving an end-to-end portability rate
of 87.5\%. The 0.85-similarity deduplication in stage~5 collapses these
656 variants to \textbf{441}, which aggregate into \textbf{155
canonical technique identifiers} (median 2 variants per ID, long tail
to 16). Full per-stage survival numbers are in Appendix~\ref{app:funnel}.

\subsection{Coverage of the Unix-competence surface}
\label{sec:coverage}

Post-hoc taxonomy assigns the 155 IDs to sixteen families spanning
filesystem metadata, binary and ELF internals, the shell environment,
system services, shell-level cryptographic primitives, processes and
IPC, logs, text-processing tools, databases and serialization formats,
media and network formats, and encoding and steganographic devices
(Figure~\ref{fig:library}a). Configuration files ($\sim$22 IDs), text
tools ($\sim$22), and text encoding and ciphers ($\sim$16) are the
largest general-purpose Unix surfaces. The smaller families are the
distinctive ones the library is designed to exercise: filesystem
metadata ($\sim$12), processes and IPC ($\sim$12), binary/ELF
($\sim$8), and crypto/cert/token formats ($\sim$9). Every canonical
entry has a canonical solution of one to five shell commands invoking an
OS- or format-specific feature. Six worked exemplars (with
\texttt{plant.sh}/\texttt{recovery.sh} pairs) are in
Appendix~\ref{app:exemplars}; two end-to-end solver trajectories
illustrating the agent behavior the surface elicits are in
Appendix~\ref{app:trajectories}; the full list of 155 identifiers is in
Appendix~\ref{app:tech-list}; the family table with representative IDs
is in Appendix~\ref{app:family}.

\section{Environment generation}
\label{sec:pipeline}

\subsection{Multi-flag containers}
\label{sec:multiflag}

Given a list of eight techniques and a seed, the environment generator
instantiates a pre-built \texttt{ctf-base} container (Ubuntu plus
standard tools). It can also dress the filesystem as one of seven
plausible server roles: \texttt{webserver}, \texttt{database},
\texttt{devbox}, \texttt{cicd}, \texttt{mailserver},
\texttt{monitoring}, or \texttt{gateway}. Each role supplies
role-appropriate users, hostnames, service configurations, fictitious
logs, and shell histories from templates; for example, an agent in the
\texttt{webserver} role sees \texttt{www-data}'s crontab, nginx
configuration, and SSL certificates. For each of the eight techniques,
the generator samples a target directory from \texttt{/home/user/},
\texttt{/var/tmp/}, \texttt{/opt/app/}, \texttt{/srv/}, and
\texttt{/tmp/}, generates a fresh flag, base64-encodes the technique's
\texttt{plant.sh} into the container, runs it with the chosen target
directory and flag, and records its exit status and output. The plant
script is then deleted, so it cannot be recovered from disk. The routine
returns a \texttt{MultiFlagEnv} containing the container name, seed, and
manifest of planted flags.

We plant eight flags per container rather than one. With a single
flag, per-rollout reward is binary, the policy receives signal only on
the (initially rare) successful rollouts, and group-relative methods
like GRPO lose most batches to constant-reward filtering. Spreading
reward across multiple sub-tasks per episode is a standard
densification move in sparse-reward RL \citep{andrychowicz2017her,
ng1999reward}, and we adopt it as a practical training choice for
validating the library under RL rather than a claim about the optimal
$n_{\text{flags}}$. The trade-off is that each container consumes
eight techniques rather than one, which the pool turnover below
partially compensates for.

\paragraph{Reward.} Per-rollout reward is
$\textit{points\_earned} - \textit{turn\_cost} \times
\textit{turns\_used}$, with 
$\textit{turn\_cost}=1$ (a maximum penalty
of about $-18$) and per-technique points defaulting to fifteen. Each
incorrect flag submission costs one point. We do not clip reward at
zero; negative-reward rollouts still provide within-group gradient
signal. The shape encodes two objectives directly (find the flag,
and find it without burning the turn budget) and was chosen, not
swept; we leave reward-shape sensitivity to future work.\looseness=-1

\subsection{Sticky-pool skill scheduling}
\label{sec:stickypool}

The scheduler maintains a rotating pool of fifty techniques from the
library. Each environment samples eight techniques uniformly from the
current pool. After each batch, \texttt{\_rotate\_pool} drops the top
$\textit{rotation\_rate} \times \textit{pool\_size}=15$ pool members by
per-technique solve rate, so mastered techniques graduate out, plus one
guaranteed random removal for exploration. Replacements come from the
$\sim$105 techniques currently outside the pool, preferring those with
the fewest prior attempts. Each batch exposes the model to
$8 \text{ groups} \times 8 \text{ flags}=64$ technique-instances, and
each rotation exchanges fifteen techniques. Across the forty-batch runs
we report, the scheduler cycles most of the 155-technique library
through the pool at least once.

\section{Evaluation protocol}
\label{sec:eval}

\subsection{Training as a probe}
\label{sec:training}

We fine-tune Qwen3-8B (\texttt{Qwen/Qwen3-8B})
\citep{yang2025qwen3} with rank-32 LoRA adapters. Training is used as a
probe of the generated surface, not as a claim that this recipe is an
optimal RL pipeline.

\paragraph{Optional SFT format pass.} Before GRPO, we optionally run a
brief supervised pass on 492 successful solver trajectories from an
earlier Haiku-solver run against the library. The purpose is to align
the model's output style and tool-call format with the unix-ctf harness
($\sim$92 gradient updates; full
hyperparameters in Appendix~\ref{app:hyperparams}). We treat this stage as harness adaptation
rather than capability transfer; one of the two GRPO initializations
skips it entirely. The empirical claims concern
the \emph{environment's} pedagogical content: whether GRPO on the
unix-ctf surface produces measurable transfer. They are not claims about
the intrinsic properties of Qwen3-8B. We use Qwen3-8B because it
produces well-formed tool calls under our harness's tool specification
without adaptation. Preliminary attempts with Llama-3.1-8B-Instruct,
Kimi-K2.5, and Nemotron-3-Super-120B-A12B exposed harness-renderer
compatibility issues: their native tool-call serializations did not
round-trip cleanly. We therefore treat model selection as a precondition
for measurement rather than a variable to ablate. The SFT format pass
extends the same harness-alignment discipline for one of the two GRPO
initializations.

\paragraph{GRPO.} From either the SFT-formatted checkpoint or the
unmodified base model, we run forty batches of group-relative policy
optimization with 8 groups $\times$ 8 rollouts (64 trajectories per
batch), up to 18 turns per rollout, temperature 1.0, KL penalty
coefficient 0, and \texttt{remove\_constant\_reward\_groups} enabled.
Fifteen technique families spanning file encoding, filesystem,
archives, databases, logs and system state, user accounts, text
processing, checksums, and process control are held out of the training
pool via the technique-blacklist mechanism (\S\ref{sec:holdout}); the
full list appears in Appendix~\ref{app:n30}. Each batch runs eight
concurrent Docker containers on a single \texttt{n2-highmem-16} GCP
instance.

\subsection{Variants and benchmarks}
\label{sec:variants}

We compare three variants: \textbf{Base Qwen3-8B} (untrained),
\textbf{GRPO from base init} (40 batches of GRPO from the unmodified
base), and \textbf{GRPO from SFT init} (40 batches of GRPO from the
SFT-formatted checkpoint). The two GRPO runs share hyperparameters and
training-budget settings but use independent random orderings drawn by
the sticky-pool scheduler (\S\ref{sec:stickypool}); the principal
intentional difference is the brief format-adaptation pass that
precedes one of them. All evaluations use the \texttt{final} checkpoint.\looseness=-1

\paragraph{Benchmark 1: InterCode-CTF (100 tasks).} We evaluate on the
100-task picoCTF-2019 split under our own driver
(\texttt{scripts/eval\_intercode.py}), which wraps each agent action in
\texttt{/bin/bash -c \dots} before passing it to InterCode's
\texttt{step}. All IC-CTF numbers we report use this wrapped driver.
Concurrency 8, 20-turn budget per task; harness notes in
Appendix~\ref{app:icctf}.

\paragraph{Benchmark 2: InterCode-Bash fs\_1 (60 tasks).} We evaluate
on the largest filesystem variant (\texttt{fs\_1}) of NL2Bash under the
InterCode harness. Reward is continuous on $[0,1]$ and combines a
filesystem-diff signal, an MD5 equality-of-changes signal, and the
TF-IDF similarity of the final stdout to the gold command's stdout. We
report both the conventional threshold (reward $\geq 0.5$) and a
stricter threshold (reward $\geq 0.8$); Appendix~\ref{app:icbash}
shows the per-component breakdown for our variants.

\paragraph{Benchmark 3: Multi-family holdout ($15$ techniques $\times$
$15$ trials, $n{=}225$).} This is our internal out-of-distribution test.
During training, we blacklist 15 technique families spanning all major
library categories: file encoding, filesystem, archives, databases, logs
and system state, user accounts, text processing, checksums, and process
control. We then evaluate on $n{=}225$ pre-built environments, with 15
fresh trials per held-out technique, one planted flag per environment,
and an 18-turn budget. No instance of any held-out technique appears in
the training pool, so solving these tasks requires generalization across
families rather than memorization within one. The full holdout list is
in Appendix~\ref{app:n30}. We report Wilson 95\% confidence intervals
alongside point estimates.

\section{Empirical validation}
\label{sec:results}

We use training as an experimental probe of the generated surface, not
as a tuned RL recipe. The principal table is in Appendix~\ref{app:main}.

\subsection{The surface is learnable}
\label{sec:holdout}

\begin{figure}[t]
  \centering
  \includegraphics[width=0.7\linewidth]{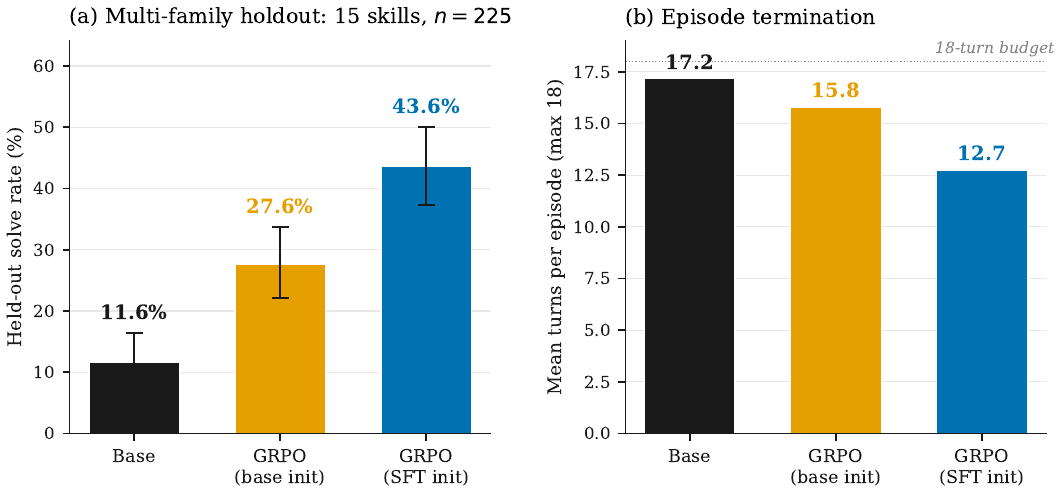}
  \caption{Multi-family holdout, $n{=}225$ single-flag environments per variant, 18-turn budget. \textbf{(a)} Aggregate solve rate with Wilson 95\% intervals. Both GRPO initializations lift substantially over base; GRPO from an SFT-initialized checkpoint reaches \GRPOSFTPCT{}, a $3.8\times$ gain over base. \textbf{(b)} Mean turns per episode (max 18). The SFT-initialized variant uses the budget more efficiently.}
  \label{fig:holdout}
\end{figure}

We fine-tune Qwen3-8B with rank-32 LoRA~\citep{hu2021lora} adapters
using GRPO~\citep{shao2024deepseekmath}, then evaluate on the same
15-skill held-out set (full list in Appendix~\ref{app:n30}).
\textbf{Base solves \BASEPCT{}; GRPO from base init reaches
\GRPOBASEPCT{}; GRPO from SFT init reaches \GRPOSFTPCT{}.} Both GRPO
initializations transfer substantially above base, and the SFT-formatted
initialization adds another \GRPOSFTDELTA{}\,pp over the base
initialization on this multi-family set. On \texttt{timeout\_race}
alone, a dedicated $n{=}30$ run gives GRPO-from-SFT-init 20/30 versus
base 8/30 (Fisher's exact $p \approx 0.004$); details are in
Appendix~\ref{app:n30}.\looseness=-1

\subsection{Transfer to InterCode-CTF}
\label{sec:redistribution}

\begin{figure}[t]
  \centering
  \includegraphics[width=0.85\linewidth]{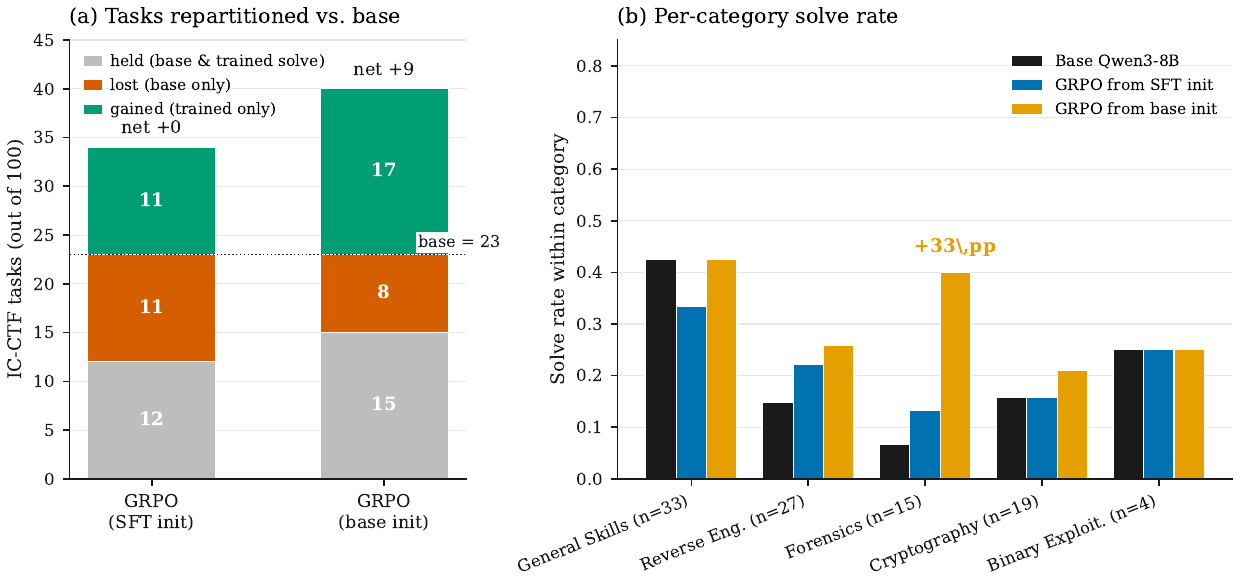}
  \caption{InterCode-CTF capability redistribution. \textbf{(a)} Each trained variant's IC-CTF-100 results decomposed against base into \textit{held} (both solve), \textit{lost} (base only), and \textit{gained} (trained only); the dotted line marks the base solve rate. \textbf{(b)} Per-category solve rate.}
  \label{fig:redistribution}
\end{figure}

On the external InterCode-CTF benchmark (100 tasks, picoCTF-2019), the
GRPO variants change \emph{which categories} the model solves more than
they change the aggregate. \textbf{The largest category-level transition
is GRPO from base init on Forensics: $1/15 \to 6/15$, a $\bf{+33}$\,pp gain}
(Figure~\ref{fig:redistribution}b). Reverse Engineering also improves.
GRPO from SFT init shows the same qualitative redistribution despite no
aggregate IC-CTF gain (Figure~\ref{fig:redistribution}a). The training
surface therefore appears to raise solve rate preferentially on the
Unix-competence-adjacent subset of IC-CTF. Full task-level decomposition
and the per-category table are in Appendix~\ref{app:cat}. A secondary RL-pipeline observation is that the two GRPO initializations
diverge between training reward and IC-CTF transfer. GRPO-from-SFT-init
peaks at $+27.0$ batch reward but reaches only 23\% on IC-CTF, while
GRPO-from-base-init never exceeds $+1.45$ and reaches 32\%. The ranking
inverts; per-batch curves and discussion are in
Appendix~\ref{app:fig-reward}.\looseness=-1

\textbf{Two GRPO initializations diverge across target distributions.}
GRPO from base init reaches \GRPOBASEPCT{} on the multi-family holdout
and 32\% on IC-CTF; GRPO from SFT init reaches \GRPOSFTPCT{} on
multi-family but only 23\% on IC-CTF. The ranking inverts. We use the
SFT pass only for harness format adaptation and do not isolate its
mechanism; practitioners should treat format-adaptation data as an
experimental variable rather than a fixed step. Each variant is
single-seed; per-skill rates are noisy ($n{\approx}15$); we do not
ablate $n_{\text{flags}}$, the scheduler, per-role rates, or non-final
checkpoints; IC-Bash \texttt{answer\_similarity} is identically zero
across variants (Appendix~\ref{app:icbash}).
\section{Conclusion and limitations}
Unix-ctf is a novel approach for generating synthetic Unix competence tasks---an eval surface that previous terminal task generation methods do not target directly. Furthermore, unix-ctf achieves a higher task yield than comparable generation methods by procedurally constructing tasks from a canonicalized library of CTF techniques. Finally, we demonstrate that the skills learned from training on unix-ctf can improve downstream performance on hand-curated IC-CTF tasks.

A limitation of unix-ctf is that it relies on LLMs to construct its technique library, so techniques poorly represented in the model during harvesting will not make it into the generated tasks. Additionally, our approach is constrained to CTF style tasks, which do not measure the full breadth of tasks that can be accomplished with unix. A promising direction for future work is to parametrise and train a unix-task generator to build curriculums of increasingly difficult unix CTF tasks.

\textbf{Dual-use and release.} The 155 techniques form a surface for
offensive-security training: each entry teaches recovery of a
Unix-resident artifact, a capability relevant to red-team agents and
defenders training against them. The library is already in active use in
such a pipeline. Releasing the 155-ID set with the associated
\texttt{plant.sh}/\texttt{recovery.sh} pairs would lower the barrier to
equivalent offensive training corpora. We therefore release enough
methodological detail for academic reproduction, including pipeline
factoring, bidirectional contract, worked exemplars, full ID
enumeration, and judge rubric, while withholding the artifact itself.

\clearpage
\bibliography{references}

\clearpage
\appendix
\begin{center}
\textbf{\Large Appendix} \\[2pt]
\textit{Supplementary material for ``unix-ctf''}
\end{center}
\vspace{1em}

\section{Principal results table}
\label{app:main}

Table~\ref{tab:main} reports the four-variant result summary referenced
by \S\ref{sec:results}.

\begin{table}[h]
\centering
\caption{Principal results. For IC-Bash we report the conventional reward $\geq 0.5$ threshold, the stricter reward $\geq 0.8$ threshold, and raw mean reward; see Appendix~\ref{app:icbash} for why the 0.5 threshold is misleading. Multi-family holdout aggregates $n{=}225$ single-flag environments uniformly sampled across the 15 evaluable held-out skills (see \S\ref{sec:holdout}).}
\label{tab:main}
\small
\setlength{\tabcolsep}{4pt}
\resizebox{\textwidth}{!}{%
\begin{tabular}{lccc}
\toprule
Variant & IC-CTF-100 & IC-Bash fs\_1 ($\geq$0.5 / $\geq$0.8 / mean) & Multi-family holdout ($n{=}225$) \\
\midrule
Base Qwen3-8B                                  & 23 / 100 (23.0\%) & 48 / 60 / \textbf{0} / 60, mean 0.536 & 26 / 225 (11.6\%) \\
GRPO from SFT init (40 batches)                & 23 / 100 (23.0\%) & 42 / 60 / \textbf{0} / 60, mean 0.469 & \textbf{98 / 225 (43.6\%)} \\
\textbf{GRPO from base init (40 batches)}      & \textbf{32 / 100 (32.0\%)} & \textbf{52} / 60 / \textbf{0} / 60, mean \textbf{0.581} & 62 / 225 (27.6\%) \\
\bottomrule
\end{tabular}%
}
\end{table}

\section{Stage-by-stage survival}
\label{app:funnel}

\begin{table}[H]
\centering
\caption{Stage-by-stage survival from 750 raw harvest attempts.}
\label{tab:funnel}
\begin{tabular}{lrr}
\toprule
Stage & Survivors & Cumulative survival \\
\midrule
Stage 1 (exploration)                            & $\sim$712 & $\sim$95\% \\
Stage 2 (mechanical check)                       & $\sim$700 & $\sim$93\% \\
Stage 3 (synthesis, Haiku format)                & $\sim$693 & $\sim$92\% \\
Stage 4 (portability validation)                 & \textbf{656} & \textbf{87.5\%} \\
Stage 5 (canonicalization, $\geq$0.85 sim. dedup) & \textbf{441} & 58.8\% \\
\bottomrule
\end{tabular}
\end{table}

\newpage
\section{Library composition by family}
\label{app:family}

The 155 canonical technique identifiers ($k=441$ deduplicated variants)
decompose into sixteen post-hoc families. Counts are approximate because
some IDs span more than one family.

\begin{table}[H]
\centering
\caption{Library composition by family (approx.\ 155 IDs; overlap allowed).}
\label{tab:family}
\small
\setlength{\tabcolsep}{4pt}
\begin{tabular}{@{}lrp{8.6cm}@{}}
\toprule
Family & \# IDs & Representative IDs \\
\midrule
Text encoding / ciphers     & $\sim$16 & \texttt{base64}, \texttt{base64\_then\_rot13}, \texttt{caesar\_shift\_n}, \texttt{atbash\_cipher}, \texttt{vigenere\_cipher}, \texttt{uuencode}, \texttt{morse\_code\_text}, \texttt{iconv\_charset} \\
Compression / archives      & $\sim$10 & \texttt{ar\_archive}, \texttt{bzip2\_file}, \texttt{cpio\_archive}, \texttt{gzip\_then\_base64}, \texttt{tar\_nested}, \texttt{xz\_file}, \texttt{zip\_password}, \texttt{zstd\_compress} \\
Filesystem metadata         & $\sim$12 & \texttt{xattr}, \texttt{acl\_metadata}, \texttt{chattr\_immutable}, \texttt{file\_capabilities}, \texttt{sticky\_setuid}, \texttt{mtime\_pre\_epoch}, \texttt{hardlink}, \texttt{symlink\_chain}, \texttt{sparse\_file}, \texttt{whiteout\_overlay} \\
Filename tricks             & $\sim$5  & \texttt{filename\_control\_chars}, \texttt{filename\_leading\_dash}, \texttt{unicode\_homoglyph}, \texttt{punycode\_idn}, \texttt{combining\_diacritics} \\
Unicode / stego             & $\sim$6  & \texttt{whitespace\_steg}, \texttt{variation\_selectors}, \texttt{tag\_characters}, \texttt{braille\_unicode} \\
Binary / ELF                & $\sim$8  & \texttt{elf\_note\_section}, \texttt{elf\_gnu\_build\_id}, \texttt{nm\_symbol\_name}, \texttt{strings\_binary}, \texttt{xxd\_hex}, \texttt{gcc\_const\_string}, \texttt{strace\_trace} \\
Config files                & $\sim$22 & \texttt{ini\_config}, \texttt{toml\_config\_file}, \texttt{sshd\_config\_match}, \texttt{sshd\_auth\_keys\_options}, \texttt{sudoers\_comment}, \texttt{systemd\_unit\_file}, \texttt{systemd\_drop\_in}, \texttt{cron\_d\_file}, \texttt{crontab\_user}, \texttt{anacrontab\_entry}, \texttt{udev\_rule}, \texttt{ldso\_conf\_d}, \texttt{pam\_config}, \texttt{nsswitch\_conf\_source} \\
Shell env                   & $\sim$5  & \texttt{bash\_alias}, \texttt{bash\_function}, \texttt{bash\_history}, \texttt{bash\_profile}, \texttt{etc\_environment\_var} \\
Crypto / certs / tokens     & $\sim$9  & \texttt{gpg\_symmetric}, \texttt{openssl\_enc}, \texttt{openssl\_dgst\_salted}, \texttt{cert\_subject}, \texttt{x509\_custom\_oid}, \texttt{htpasswd\_hash}, \texttt{jwt\_hs256\_token}, \texttt{hmac\_signed\_token} \\
Processes / IPC             & $\sim$12 & \texttt{background\_process}, \texttt{flock\_lockfile}, \texttt{named\_pipe}, \texttt{unix\_socket}, \texttt{shm\_segment}, \texttt{signal\_handler}, \texttt{timeout\_race}, \texttt{proc\_maps}, \texttt{proc\_fd}, \texttt{ss\_connections}, \texttt{iptables\_rules} \\
Logs / system               & $\sim$6  & \texttt{syslog}, \texttt{journalctl}, \texttt{dmesg\_kernel}, \texttt{auth\_log}, \texttt{mail\_spool}, \texttt{script\_replay} \\
Text tools                  & $\sim$22 & \texttt{awk\_begin\_end}, \texttt{awk\_field}, \texttt{sed\_hold\_space}, \texttt{sed\_transform}, \texttt{tac\_reverse}, \texttt{tr\_cipher}, \texttt{xargs\_exec} \\
DB / serialization          & $\sim$12 & \texttt{sqlite\_blob}, \texttt{sqlite\_query}, \texttt{python\_marshal}, \texttt{python\_pickle}, \texttt{python\_shelve}, \texttt{ruby\_yaml\_obj}, \texttt{php\_serialize}, \texttt{msgpack\_binary}, \texttt{protobuf\_serialized}, \texttt{plist\_xml\_apple} \\
Media / network formats     & $\sim$5  & \texttt{pcap\_icmp\_payload}, \texttt{flac\_vorbis\_comment}, \texttt{jpeg\_exif\_comment}, \texttt{pdf\_metadata\_info} \\
Users / OS                  & $\sim$6  & \texttt{passwd\_shadow}, \texttt{user\_creation}, \texttt{unusual\_location}, \texttt{mktemp\_tmpfile} \\
Math                        & $\sim$3  & \texttt{bc\_math}, \texttt{factor\_prime}, \texttt{find\_size} \\
\bottomrule
\end{tabular}
\end{table}

\section{Six worked library exemplars}
\label{app:exemplars}

Each technique below is extracted from
\texttt{data/library/<technique\_id>.json}, \texttt{variants[0]}. Each
script is parameterized by a target directory and a flag; none hardcodes
either value. The six examples were selected to span the families in
Table~\ref{tab:family} and to illustrate the range of operating-system
knowledge exercised by the library.

\subsection{\texttt{xattr}: storing the flag in a POSIX extended attribute on a decoy file}

\begin{lstlisting}[language=bash]
# plant.sh
echo "nothing to see here" > "$target_dir/decoy.txt"
encoded=$(echo -n "$flag" | base64)
setfattr -n user.secret -v "$encoded" "$target_dir/decoy.txt"
\end{lstlisting}

\begin{lstlisting}[language=bash]
# recovery.sh
getfattr --only-values -n user.secret "$target_dir/decoy.txt" | base64 -d
\end{lstlisting}

The decoy file prints ``nothing to see here'' under \texttt{cat}, and
standard inspection commands such as \texttt{stat}, \texttt{ls -l},
\texttt{file}, \texttt{strings}, and a recursive \texttt{grep} for the
flag prefix do not surface the flag. The data lives in the
\texttt{user.*} extended-attribute namespace, a POSIX/Linux filesystem
feature with no direct analogue in ordinary programming; the canonical
access path is \texttt{getfattr -n}, or equivalently the
\texttt{getxattr(2)} syscall. Solving the task requires knowing that
this namespace exists and identifying the attribute name to query, or
dumping all user-namespace attributes on every file. A solver who uses
Python still has to invoke \texttt{os.getxattr}, which is precisely the
Unix knowledge being tested.

\subsection{\texttt{elf\_gnu\_build\_id}: embedding the flag as the \texttt{.note.gnu.build-id} of a stock ELF}

\begin{lstlisting}[language=bash]
# plant.sh (abbreviated)
flag_desc=$(echo -n "$flag" | xxd -p)
python3 <<PYEOF
import struct
name = b"GNU\x00"
desc_raw = bytes.fromhex("$flag_desc")
desc_padded = desc_raw + b"\x00" * ((4 - len(desc_raw) % 4) % 4)
note = struct.pack("<III", len(name), len(desc_raw), 3) + name + desc_padded
with open("buildid.note", "wb") as f: f.write(note)
PYEOF
cp /bin/ls ./hidden_elf
objcopy --update-section .note.gnu.build-id=buildid.note ./hidden_elf ./hidden_elf_marked
rm -f buildid.note hidden_elf
mv hidden_elf_marked hidden_elf
\end{lstlisting}

\begin{lstlisting}[language=bash]
# recovery.sh
readelf -n "$target_dir/hidden_elf" 2>/dev/null \
    | grep -oE 'Build ID: [0-9a-f]+' | awk '{print $3}' | xxd -r -p
\end{lstlisting}

The \texttt{file} utility identifies \texttt{hidden\_elf} as a 64-bit
ELF executable, while \texttt{strings} surfaces nothing because the flag
is hex-encoded. Solving the task requires familiarity with the ELF
note-section convention and with \texttt{readelf -n}.

\subsection{\texttt{mtime\_pre\_epoch}: pre-1970 timestamp as the discriminator}

\begin{lstlisting}[language=bash]
# plant.sh: hide real file among 10 decoys, marked by pre-epoch mtime
mkdir -p "$target_dir/data"
cd "$target_dir/data"
for i in {1..10}; do echo "random log entry $i" > "log_$i.txt"; done
echo -n "$flag" | base64 > note.txt
touch -d '1969-06-15 12:00:00' note.txt
\end{lstlisting}

\begin{lstlisting}[language=bash]
# recovery.sh
find "$target_dir/data" -type f ! -newermt '1970-01-01' -exec cat {} \; | base64 -d
\end{lstlisting}

Every file is plain text, and \texttt{ls -lt} shows eleven similar
files. The discriminating signal is the negative modification time, a
filesystem attribute that most \texttt{find} invocations never exercise.

\subsection{\texttt{shm\_segment}: tmpfs storage, doubly encoded}

\begin{lstlisting}[language=bash]
# plant.sh
encoded=$(echo -n "$flag" | base64 | tr 'A-Za-z' 'N-ZA-Mn-za-m')
echo -n "$encoded" > /dev/shm/.cache_seg_42
\end{lstlisting}

\begin{lstlisting}[language=bash]
# recovery.sh
cat /dev/shm/.cache_seg_42 | tr 'N-ZA-Mn-za-m' 'A-Za-z' | base64 -d
\end{lstlisting}

The flag does not live on the persistent filesystem, and recursive
searches rooted at \texttt{/} commonly exclude \texttt{/dev}. Solving
the task requires knowing that \texttt{/dev/shm} is a tmpfs mount, then
applying ROT13 and base64 in the correct order.

\subsection{\texttt{x509\_custom\_oid}: flag stored as a UTF8String in a custom X.509 extension}

\begin{lstlisting}[language=bash]
# plant.sh
cat > openssl.cnf << 'EOFCNF'
[ req ]
distinguished_name = req_dn
x509_extensions = v3_ext
prompt = no
[ req_dn ]
CN = ctf.example.com
[ v3_ext ]
1.3.6.1.4.1.55555.1.1 = ASN1:UTF8String:PLACEHOLDER_FLAG
EOFCNF
sed -i "s/PLACEHOLDER_FLAG/$flag/g" openssl.cnf
openssl req -x509 -newkey rsa:2048 -nodes -keyout key.pem -out cert.pem -days 365 -config openssl.cnf
rm openssl.cnf
\end{lstlisting}

\begin{lstlisting}[language=bash]
# recovery.sh
openssl x509 -in "$target_dir/cert.pem" -text -noout 2>/dev/null | grep -oE 'flag\{[^}]+\}'
\end{lstlisting}

Inspecting \texttt{cert.pem} as text reveals only a PEM block.
\texttt{openssl x509 -text} is the specific command that decodes the
ASN.1 custom object identifier, and most agents will not invoke it
without reasoning about public-key infrastructure.

\subsection{\texttt{whiteout\_overlay}: one character per directory name, separated by zero-width spaces}

\begin{lstlisting}[language=bash]
# plant.sh
ZWSP=$'\xe2\x80\x8b'   # U+200B
mkdir -p "$target_dir/hidden"
cd "$target_dir/hidden"
for (( idx=0; idx<${#flag}; idx++ )); do
  ch="${flag:$idx:1}"
  printf -v padded "%02d" $idx
  mkdir -- "data_${padded}_${ZWSP}${ch}${ZWSP}end"
done
\end{lstlisting}

\begin{lstlisting}[language=bash]
# recovery.sh
cd "$target_dir/hidden"
ls -1 | sort | python3 -c "import sys; print(''.join(l.rstrip('\n').split('_',2)[2].replace(chr(0x200b),'')[0] for l in sys.stdin))"
\end{lstlisting}

A directory listing shows about twenty-two directories with names of the
form \texttt{data\_00\_*end}, \texttt{data\_01\_*end}, and so on.
Zero-width spaces are invisible in most terminals, so solving the task
requires noticing that the names are not actually identical, sorting
them lexically, and stripping the zero-width spaces.

\section{Worked solver trajectories}
\label{app:trajectories}

To make the kind of agent behavior the library elicits concrete, we reproduce two successful solver trajectories from a Claude Haiku 4.5 reference run. Lines beginning with \texttt{>} are the agent's narration before each tool call; lines beginning with \texttt{\$} are the bash invocations the agent issued; following lines are the captured stdout. Tool calls and tool returns are abridged for space; the full message logs are recorded in the project artifact directory.

\paragraph{Exemplar 1: \texttt{passwd\_shadow} (3 turns).} The instruction primes the agent for ``a system administrator account [with] hidden information embedded in its metadata.'' The agent recognizes the GECOS convention of \texttt{/etc/passwd} (the comma-separated metadata in the fifth field) and extracts a base64-encoded flag.

\begin{lstlisting}[basicstyle=\ttfamily\footnotesize, breaklines=true, columns=fullflexible, xleftmargin=0pt]
[turn 1]
> I'll start by exploring the system to find the hidden information in the
> administrator account metadata.
$ whoami
user
$ cat /etc/passwd | grep -E "root|admin"
root:x:0:0:root:/root:/bin/bash
gnats:x:41:41:Gnats Bug-Reporting System (admin):/var/lib/gnats:/usr/sbin/nologin
sysadmin2024:x:1001:1001:System Admin,ZmxhZ3syM2FiZGM0NGI5ODliMjJjfQ==,Room 101,555-1234:/home/sysadmin2024:/bin/false
$ ls -la /home
drwxr-x--- 1 sysadmin2024 sysadmin2024 54 Apr  8 20:22 sysadmin2024
drwxr-x--- 1 user         user         54 Apr  6 22:14 user

[turn 2]
> Excellent! I found something interesting. In /etc/passwd, the user
> sysadmin2024 has what looks like base64-encoded text in the GECOS field
> (the comment field): ZmxhZ3syM2FiZGM0NGI5ODliMjJjfQ==. Let me decode it.
$ echo "ZmxhZ3syM2FiZGM0NGI5ODliMjJjfQ==" | base64 -d
flag{23abdc44b989b22c}

[turn 3]
> Perfect -- the flag was hidden in the GECOS field of the system
> administrator account. The encoded string decoded to reveal the flag.
submit_flag(flag{23abdc44b989b22c})  ->  Acknowledged.
\end{lstlisting}

The Unix-specific knowledge here is the structure of \texttt{/etc/passwd} (seven colon-separated fields, with the fifth carrying free-form GECOS metadata as comma-separated subfields) and the convention of stashing administrative metadata there rather than in a separate database. A solver that defaults to grepping for \texttt{flag\{} would never see this.

\paragraph{Exemplar 2: \texttt{mail\_spool} (4 turns).} The instruction primes for ``a secret message [\dots] hidden within the system's mail infrastructure'' and ``a custom email header containing encoded data.'' The agent probes the standard mail-spool locations (\texttt{/var/mail/}, \texttt{/var/spool/mail/}, \texttt{\textasciitilde/Mail/}) and recognizes the mailbox file format with custom \texttt{X-} headers.

\begin{lstlisting}[basicstyle=\ttfamily\footnotesize, breaklines=true, columns=fullflexible, xleftmargin=0pt]
[turn 1]
> I'll help you find the flag hidden in the mail infrastructure. Let me
> start by exploring the filesystem structure and looking for mail files.
$ ls -la /var/mail/
-rw-r--r-- 1 root mail 576 Apr  8 20:20 testuser
$ ls -la /var/spool/mail/
-rw-r--r-- 1 root mail 576 Apr  8 20:20 testuser
$ ls -la ~/Mail/ 2>/dev/null || ls -la ~/mail/ 2>/dev/null || echo "No Mail directory"
No Mail directory

[turn 2]
> Great! I found a mailbox file at /var/mail/testuser. Let me examine its
> contents.
$ cat /var/mail/testuser
From sender@example.com  Mon Mar 22 14:30:00 2024
Return-Path: <sender@example.com>
Received: from mail.example.com (mail.example.com [192.168.1.1])
        by localhost with SMTP id ABC123
        for <testuser@localhost>; Mon, 22 Mar 2024 14:30:00 +0000
From: Admin <sender@example.com>
To: testuser@localhost
Date: Mon, 22 Mar 2024 14:30:00 +0000
Subject: System Notification
Message-ID: <msg001@example.com>
X-Secret-Code: ZmxhZ3s5NjI3MGQ2NWJhZDdhMTNjfQ==
Content-Type: text/plain; charset=UTF-8
Content-Length: 42

This is an automated system notification.
Please ignore this message.

[turn 3]
> Perfect! I found a custom email header called X-Secret-Code with what
> appears to be base64-encoded data. Let me decode it.
$ echo "ZmxhZ3s5NjI3MGQ2NWJhZDdhMTNjfQ==" | base64 -d
flag{96270d65bad7a13c}

[turn 4]
> Excellent! I found the flag. Let me submit it.
submit_flag(flag{96270d65bad7a13c})  ->  Acknowledged.
\end{lstlisting}

The Unix-specific knowledge is twofold: the mail-spool location convention (\texttt{/var/mail/<user>}) and the mailbox file format with \texttt{From\_} envelope lines and arbitrary RFC-822 \texttt{X-}-prefixed headers. Both are instances of the broader ``concealed data inside a structured system file'' pattern that the library exercises across $\sim$22 configuration-file techniques and $\sim$12 filesystem-metadata techniques (\S\ref{sec:coverage}).

\section{InterCode-CTF capability redistribution: reference data}
\label{app:fig-ctf}

This section gives the reference data behind \S\ref{sec:redistribution}
and Figure~\ref{fig:redistribution}.

\subsection{Per-category solve counts}
\label{app:cat}

\begin{table}[h]
\centering
\caption{IC-CTF per-category decomposition (count / category-N). Category labels are reconstructed canonically by \texttt{task\_id} from the full InterCode-CTF task list, so every variant is evaluated against the same fixed 100-task partition. Cell totals correspond to the principal table (Table~\ref{tab:main}).}
\label{tab:cat}
\small
\begin{tabular}{lccc}
\toprule
Category & Base & GRPO from SFT init & GRPO from base init \\
\midrule
General Skills        & 14 / 33 (42\%) & 11 / 33 (33\%) & \textbf{14 / 33 (42\%)} \\
Reverse Engineering   & 4 / 27 (15\%)  & 6 / 27 (22\%)  & \textbf{7 / 27 (26\%)} \\
Forensics             & 1 / 15 (7\%)   & 2 / 15 (13\%)  & \textbf{6 / 15 (40\%)} \\
Cryptography          & 3 / 17 (18\%)  & 3 / 19 (16\%)  & \textbf{4 / 19 (21\%)} \\
Binary Exploitation   & 1 / 4 (25\%)   & 1 / 4 (25\%)   & 1 / 4 (25\%) \\
Web Exploitation      & 0 / 2          & 0 / 2          & 0 / 2 \\
\bottomrule
\end{tabular}
\end{table}

\subsection{Example tasks by class}

The puzzle-versus-programming-task distinction is qualitative. The
following are representative examples.
\begin{itemize}
\item \textbf{Puzzle (forfeited by training)}: \texttt{picoCTF\{61\}} (concatenating a file whose contents are the number 61); \texttt{picoCTF\{CRYPTOISFUN\}} (a trivial Caesar shift on a single-line input); several General Skills tasks with single-line gold flags.
\item \textbf{Programming-style (acquired by training)}: \texttt{picoCTF\{0x6e8\}} (staged reverse engineering across multiple related binaries); \texttt{picoCTF\{p4ck37\_5h4rk\_\ldots\}} (pcap filtering and payload extraction); \texttt{picoCTF\{not\_too\_bad\_of\_a\_problem\}} (multi-step chained cryptography).
\end{itemize}

\subsection{Task-level outcome decomposition}
\begin{itemize}
\item \textbf{GRPO-from-base versus base (net $+9$)}: both 15; base-only 8 (6 General Skills, 2 Reverse Engineering); GRPO-only 17 (5 Forensics, 6 General Skills, 5 Reverse Engineering, 1 Cryptography); neither 60.
\item \textbf{GRPO-from-SFT versus base (net $\pm 0$)}: both 12; base-only 11 (predominantly trivial General Skills decodings); GRPO-only 11 (more challenging Reverse Engineering, Cryptography, Forensics); neither 66.
\end{itemize}

\subsection{Why Forensics specifically}
The transition from $1/15$ ($7\%$) for base to $6/15$ ($40\%$) for
GRPO-from-base is the largest category-level move in our sweep. By the
definition in \S\ref{sec:intro}, Forensics is the IC-CTF category most
aligned with the library's design intent: it rewards systematic
filesystem exploration and parsing of non-obvious artifacts, including
pcaps, carved images, filesystem metadata, and steganographically
concealed data. The $+33$\,pp improvement is consistent with the
training distribution teaching these skills, and is our most direct
evidence that the library's coverage transfers to a third-party
benchmark.

\section{GRPO training-reward curves}
\label{app:fig-reward}

\begin{figure}[H]
  \centering
  \includegraphics[width=0.7\linewidth]{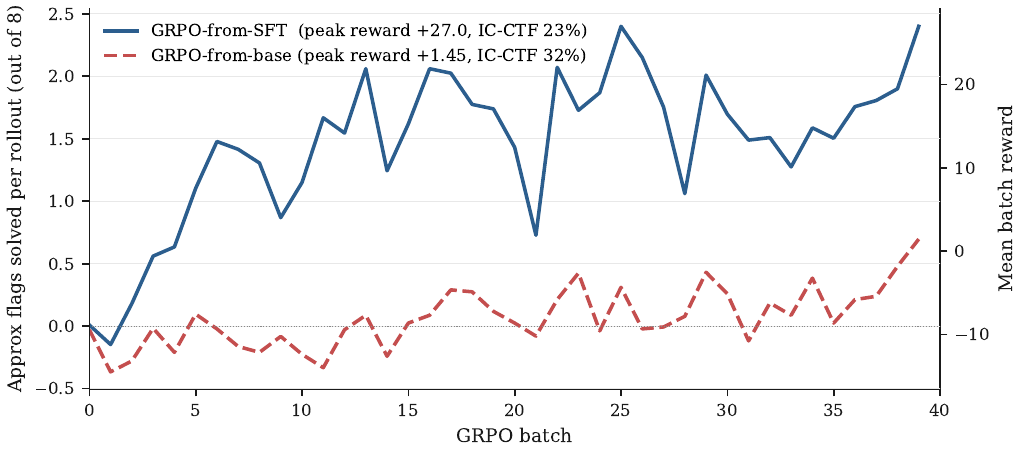}
  \caption{Per-batch GRPO training reward over forty batches. Solid blue: GRPO-from-SFT (peaks at $+27.0$). Dashed red: GRPO-from-base (peaks at $+1.45$). Despite the training-reward gap, GRPO-from-base reaches 32\% on InterCode-CTF and GRPO-from-SFT only 23\%; the variant ranking inverts.}
  \label{fig:decoupling}
\end{figure}

GRPO-from-SFT climbs to roughly 2.4 flags per rollout while
GRPO-from-base barely escapes zero (Figure~\ref{fig:decoupling}), yet
GRPO-from-base is the better InterCode-CTF solver (32\% vs.\ 23\%).
The SFT warm-start helps when evaluation matches the training
distribution (\S\ref{sec:holdout}) but hurts when the distributions
diverge (\S\ref{sec:redistribution}). Pipelines targeting OOD transfer
should therefore monitor downstream evaluations directly rather than
using training reward as a proxy.

\section{InterCode-CTF harness notes}
\label{app:icctf}

Our IC-CTF eval driver
(\texttt{scripts/eval\_intercode.py}) prepends \texttt{/bin/bash -c
\dots} to every agent action before passing it to the
\texttt{CTFEnv.step} method, with a single exception for bare
\texttt{cd <path>} commands so that \texttt{CTFEnv}'s workdir-tracking
branch persists directory changes across calls. We document this so
that comparisons across IC-CTF results from different driver
implementations can be calibrated. We also hint to the agent in the
system prompt that task files reside under a per-task subdirectory,
which avoids unnecessary path-discovery turns in IC-CTF's shared-image
layout.

\section{Training hyperparameters}
\label{app:hyperparams}

\textbf{Base model.} Qwen3-8B (\texttt{Qwen/Qwen3-8B})
\citep{yang2025qwen3} with rank-32 LoRA \citep{hu2021lora} adapters.
\textbf{SFT format pass.} 492 successful Haiku-solver trajectories
against the unix-ctf library, $\sim$92 gradient updates on hosted
Tinker; we use Tinker's recipe defaults for optimizer, learning rate,
and batch size at this stage and treat the pass as harness alignment
rather than capability transfer. \textbf{GRPO.} 40 batches, 8 groups
$\times$ 8 rollouts (64 trajectories per batch), maximum 18 turns per
rollout, sampling temperature 1.0, KL penalty coefficient 0,
\texttt{remove\_constant\_reward\_groups} enabled. Reward per rollout
is $\textit{points\_earned} - \textit{turn\_cost} \times
\textit{turns\_used}$, with $\textit{turn\_cost}=1$, per-technique
points 15, and $-1$ per incorrect flag submission. Each batch runs
eight concurrent Docker containers on a single
\texttt{n2-highmem-16} GCP instance. \textbf{Seeds.} Each variant is
trained with a single random seed; we have not run multi-seed
replications.

\section{InterCode-Bash per-component breakdown}
\label{app:icbash}

InterCode-Bash reward is
$\frac{1}{3}(\textit{file\_diff} + \textit{file\_changes} +
\textit{answer\_similarity})$. We report numbers in
Table~\ref{tab:main} at both the conventional ($\geq 0.5$) and stricter
($\geq 0.8$) thresholds. Disaggregating the base Qwen3-8B run on
\texttt{fs\_1} ($n{=}60$): \texttt{file\_diff} mean 0.264,
\texttt{file\_changes} mean 0.264, \texttt{answer\_similarity} mean
\textbf{0.000} (0/60 tasks above 0.05), strict success ($\geq 0.8$) of
\textbf{0/60}. The same pattern holds across our four variants:
\texttt{answer\_similarity} is zero on every task, and no variant
clears the strict threshold. We therefore use the strict threshold as
our primary IC-Bash signal in this paper and report the conventional
threshold for cross-paper comparability.

\section{Extended holdout at $n{=}30$}
\label{app:n30}

\paragraph{The 15 evaluable held-out skills.} \texttt{ar\_archive}, \texttt{bash\_history}, \texttt{crontab\_user}, \texttt{dd\_extract}, \texttt{find\_inode}, \texttt{gpg\_symmetric}, \texttt{ini\_config}, \texttt{mail\_spool}, \texttt{od\_octal}, \texttt{paste\_merge}, \texttt{sha256\_verify}, \texttt{sparse\_file}, \texttt{timeout\_race}, \texttt{xml\_grep}, \texttt{zstd\_compress}. They span archives, shell environment, configuration, filesystem metadata, IPC, logs, encoding, text-processing, checksums, and binary formats.

\S\ref{sec:holdout} gives the principal multi-family result at
$n{=}225$. As a complementary deep dive on one held-out family, we ran
an extended \texttt{timeout\_race}-only evaluation at $n{=}30$; the
corresponding table is reproduced here. Per-environment outcomes are
recorded in
\texttt{docs/paper/data/holdout\_results\_all\_variants\_n30.csv}.

\begin{table}[h]
\centering
\small
\caption{Held-out \texttt{timeout\_race} solve rate at $n=30$ fresh environments per variant. Wilson 95\% CI shown. Fisher's exact two-sided $p$ tests Base versus GRPO-from-SFT.}
\label{tab:n30}
\begin{tabular}{lccc}
\toprule
Variant & Solved & Solve rate (95\% CI) & Mean turns to solve \\
\midrule
Base Qwen3-8B          & 8 / 30           & 26.7\% \,([14.2\%, 44.4\%]) & 10.5 \\
GRPO-from-SFT          & \textbf{20 / 30} & \textbf{66.7\%} \,([48.8\%, 80.8\%]) & 5.0  \\
\bottomrule
\end{tabular}
\end{table}

Fisher's exact two-sided test gives $p \approx 0.0040$. The qualitative
finding is preserved: GRPO-from-SFT solves the held-out family at a
substantially higher rate than the base model, with non-overlapping
Wilson 95\% intervals. The mean-turns-to-solve gap is also preserved
(5.0 turns vs.\ 10.5 turns), a pattern consistent with a policy that
recognizes the family rather than discovering the flag by chance.

\paragraph{The principal ratio drops from $\sim 5\times$ to
$\sim 2.5\times$.} \emph{This revision concerns only the
\texttt{timeout\_race} solve-rate multiplier, not the unrelated
production-yield $5\times$ of \S\ref{sec:contract} (87.5\% vs.\
17.5\%), which uses different numerators and denominators.} At $n{=}8$,
the base \texttt{timeout\_race} solve rate was $1/8=12.5\%$, making the
GRPO-from-SFT result appear as a $5\times$ improvement. With the
$n{=}30$ base estimate ($8/30=26.7\%$), the ratio drops to
$\sim 2.5\times$. The absolute gap ($+40$\,pp) and statistical
significance ($p \approx 0.004$) are robust; the multiplicative
``$5\times$'' framing was an artifact of the small-$n$ base estimate. We
therefore recommend the absolute improvement, Wilson interval, and
Fisher's exact $p$ for future executions of this comparison.

\section{Task-quality audit rubric}
\label{app:audit_rubric}

Three LLM judges, \texttt{gpt-5.4-nano}
\citep{openai2026gpt54nano}, \texttt{grok-4-20-reasoning}, and
\texttt{Kimi-K2.5}, all served via Azure Foundry, are independently
given the rubric below, the task instruction, the canonical solution,
and the test specification. Each judge outputs a JSON object with one
(score, justification) cell per dimension. Aggregation is the mean of
dimension means; the headline numbers in Table~\ref{tab:audit} are
means $\pm$ standard deviation across the three judges. Per-judge
dimension means are reproduced in Table~\ref{tab:audit_per_judge}.
Grading robustness was originally a sixth dimension, but the LLM judges
could not reliably distinguish exact-match against a harness-generated
oracle from exact-match against a hardcoded LLM-authored expected string
without rubric language that prejudged the corpus. We therefore drop
that dimension and rely on a mechanical count of \texttt{assertEqual}
calls instead (99\% of Endless Terminals tasks; 0\% of unix-ctf, by construction).

\begin{table}[H]
\centering
\caption{LLM-rubric task-quality scores broken out by judge model (1 worst, 5 best; corresponds to the 3-judge mean reported in Table~\ref{tab:audit}). Each cell is the mean rubric score over all tasks in the listed corpus, scored by the listed LLM judge against the rubric defined below. Aggregate is the mean of the five non-grading-robustness dimensions. Pairwise Pearson $r$ on per-task aggregate scores: \texttt{gpt}$\times$\texttt{grok} $=0.82$, \texttt{gpt}$\times$\texttt{kimi} $=0.83$, \texttt{grok}$\times$\texttt{kimi} $=0.93$ ($n{=}559$ common tasks).}
\label{tab:audit_per_judge}
\small
\setlength{\tabcolsep}{4pt}
\resizebox{\linewidth}{!}{%
\begin{tabular}{l ccc ccc}
\toprule
 & \multicolumn{3}{c}{unix-ctf ($n{=}439$)} & \multicolumn{3}{c}{Endless Terminals ($n{=}120$)} \\
\cmidrule(lr){2-4}\cmidrule(lr){5-7}
Dimension & \texttt{gpt-5.4-nano} & \texttt{grok-4-20} & \texttt{Kimi-K2.5} & \texttt{gpt-5.4-nano} & \texttt{grok-4-20} & \texttt{Kimi-K2.5} \\
\midrule
Input richness          & 4.21 & 5.00 & 4.99 & 1.48 & 2.12 & 1.82 \\
Solution non-triviality & 3.52 & 3.75 & 3.08 & 2.33 & 3.07 & 2.23 \\
Format independence     & 3.26 & 5.00 & 5.00 & 1.10 & 1.02 & 1.08 \\
Tool surface diversity  & 2.51 & 4.11 & 3.70 & 2.09 & 3.65 & 2.80 \\
Discoverability         & 3.16 & 3.00 & 2.99 & 1.14 & 1.19 & 1.03 \\
\midrule
\textbf{Aggregate}      & \textbf{3.34} & \textbf{4.17} & \textbf{3.95} & 1.63 & 2.21 & 1.79 \\
\bottomrule
\end{tabular}
}
\end{table}

\paragraph{LLM-judge rubric (verbatim).} The text below is the exact
rubric supplied to each of the three LLM judges (\texttt{gpt-5.4-nano},
\texttt{grok-4-20-reasoning}, \texttt{Kimi-K2.5}) along with the task
instruction, canonical solution, and grading specification; the judge
returns a JSON object with one (score, justification) pair per
dimension, and per-task scores are aggregated as the mean of dimension
means.

\begin{lstlisting}[basicstyle=\ttfamily\footnotesize, breaklines=true, columns=fullflexible, xleftmargin=0pt]
Dimension 1 -- Input richness:
  5 = realistic container/system state (real logs, multiple processes, role-dressing
      with hundreds of records or a multi-file fixture)
  3 = moderately-sized fixture (>50 lines, multiple files)
  1 = tiny synthetic fixture (<=15 lines via heredoc/echo, single file)

Dimension 2 -- Solution non-triviality (depth of OS knowledge required, not solution length):
  5 = requires non-obvious OS or format knowledge (xattrs, ELF sections/notes,
      sparse files, IPC, signals, X.509 internals, /proc/* internals,
      archive metadata, kernel interfaces). Solution length is irrelevant.
  3 = requires composing familiar Unix tools in a non-trivial way
  1 = single one-line awk/grep/sed pipeline applying a stock idiom, OR trivially
      solvable in any general-purpose language with no Unix-specific knowledge

Dimension 3 -- Format independence:
  5 = minimal instructions; success depends on functional correctness
  3 = instructions specify a few output format details
  1 = instructions are heavy with format trivia (specific decimal places, comma-separated,
      sorted ascending, exact column header text) AND grading checks every detail

Dimension 4 -- Tool surface diversity:
  5 = solution uses >=2 features beyond the awk/sort/grep/sed/find/cat core
      (signals, IPC, fs metadata, binary formats, encodings, X.509, archives,
      kernel interfaces)
  3 = solution uses 2-3 of the core utilities in combination
  1 = single application of awk/grep/sed/find/sort

Dimension 5 -- Discoverability (investigation required):
  5 = task requires genuine investigation; the relevant feature is not announced
      in the prompt and the agent must explore filesystem/process/state to
      find clues
  3 = task hints at the surface but agent still must locate the specific entry point
  1 = task tells the agent exactly what tool/file/command to use ("use sqlite3 to
      query table X and write to file Y in this format")
\end{lstlisting}

\section{Skill-category classification rubric (Figure~\ref{fig:taxonomy})}
\label{app:taxonomy_rubric}

The same judge model (\texttt{gpt-5.4-nano}
\citep{openai2026gpt54nano}) classifies each task into one of four
skill categories. The per-task input is the instruction text, canonical
solution, and grading specification. The output is
\texttt{\{"category": "<one of four>", "reason": "<one sentence>"\}}.

\begin{lstlisting}[basicstyle=\ttfamily\footnotesize, breaklines=true, columns=fullflexible, xleftmargin=0pt]
Discriminating question for unix_competence vs the others: "Could a competent
Python programmer who knew nothing about Unix internals beyond subprocess.run
solve this task by reading the prompt and writing Python?" If yes, NOT
unix_competence.

CATEGORY: unix_competence
The canonical solution invokes a NON-OBVIOUS OS-, kernel-, or format-internal
feature that most engineers would not know to look for without being told. The
Python-stdlib path either does not exist or requires invoking the same OS
feature via syscall/ctypes.
Examples: getfattr, readelf -n, debugfs/find -inum, systemctl in non-trivial
ways, git plumbing (cat-file, pack files, reflog format), POSIX IPC (named
pipes, signals, fifos), filesystem metadata (sparse files, ACLs, capabilities),
pre-epoch mtimes, X.509 custom OIDs, kernel /proc internals.
NOT unix_competence: sqlite3, tar, gzip, awk, sed, find, git status/log/commit,
mkdir, chown, chmod, formatted-text generation, CSV/JSON manipulation -- these
all have clean Python equivalents.

CATEGORY: shell_flavored_coding
Glue pipeline of core text utilities (find, grep, awk, sed, sort, cut, tr, head,
tail, uniq) and/or familiar binaries (sqlite3, tar, gzip, jq, curl) for which
a Python-stdlib (or pandas/sqlite3 module) equivalent of comparable length exists.
Shell is convenient but not load-bearing.

CATEGORY: general_programming
Deliverable is a non-trivial program; success depends on programming proficiency
in a general-purpose language. Multi-hundred-line solve scripts, custom
algorithms, ML training, parsers/interpreters, network services.

CATEGORY: other
Not primarily about shell, OS features, or programming. Pure cryptographic
puzzles solvable with paper-and-pencil, web exploitation, single-string decoding,
trivia.
\end{lstlisting}

\paragraph{Per-benchmark provenance.}
\begin{itemize}\setlength\itemsep{1pt}
  \item \textbf{unix-ctf}: 155 canonical techniques (one variant per technique ID).
  \item \textbf{Endless Terminals}: 120 LLM-output task directories from our reproduction (excluding 16 \texttt{verified\_good/}+\texttt{validated\_batch1/} re-run duplicates and 19 pre-\texttt{generate\_remaining.sh} tasks in the \texttt{log analysis} and \texttt{text processing} categories whose run summaries were overwritten).
  \item \textbf{Terminal-Bench~2.0}: all 241 tasks under \texttt{original-tasks/} on the public \texttt{laude-institute/terminal-bench} repository (head of \texttt{main}).
  \item \textbf{InterCode-CTF}: \texttt{data/ctf/ic\_ctf.json} (100 tasks; picoCTF-2019).
  \item \textbf{InterCode-Bash}: \texttt{data/nl2bash/nl2bash\_fs\_\{1,2,3,4\}.json} (200 tasks).
  \item \textbf{Cybench}: 43 \texttt{metadata.json} files under \texttt{benchmark/} on \texttt{andyzorigin/cybench} (paper claims 40; we audit all 43 present).
  \item \textbf{NYU CTF Bench}: 200-task \texttt{test\_dataset.json} from \texttt{NYU-LLM-CTF/NYU\_CTF\_Bench}.
  \item \textbf{AgentBench-OS}: 26-task \texttt{dev.json} from \texttt{THUDM/AgentBench/data/os\_interaction/}.
  \item \textbf{Nemotron-Terminal}: 135-task subsample, drawn as 15 per category $\times$ 9 \texttt{skill\_based/easy} categories from the public HuggingFace dataset \texttt{nvidia/Nemotron-Terminal-Corpus} (random seed 42).
\end{itemize}

\section{Complete list of 155 canonical technique IDs}
\label{app:tech-list}

Table~\ref{tab:appendix-d} lists every canonical technique identifier in
the library, together with the number of deduplicated variants ($k$) per
identifier after the $\geq 0.85$ \texttt{difflib.SequenceMatcher}
similarity deduplication of \S\ref{sec:harvest}. Each identifier
corresponds to a JSON entry of the form
\texttt{data/library/<id>.json}, which bundles the per-variant
\texttt{plant.sh} and \texttt{recovery.sh} pair together with the
per-stage validation flags.

\begin{table}[h]
\centering
\scriptsize
\caption{All 155 canonical technique IDs in the library, with per-ID variant counts after the $\geq 0.85$ similarity dedup of \S3.1 (total 441 variants). Pulled directly from \texttt{data/library/\_manifest.json}.}
\label{tab:appendix-d}
\begin{tabular}{@{}ll@{\hskip 1.5em}ll@{\hskip 1.5em}ll@{}}
\toprule
id & $k$ & id & $k$ & id & $k$ \\
\midrule
\texttt{acl\_metadata} & 3 & \texttt{fmt\_reflow} & 6 & \texttt{seq\_pattern} & 1 \\
\texttt{anacrontab\_entry} & 1 & \texttt{gpg\_symmetric} & 2 & \texttt{services\_file\_port} & 1 \\
\texttt{apt\_pref} & 2 & \texttt{gzip\_file} & 2 & \texttt{sha256\_verify} & 1 \\
\texttt{ar\_archive} & 3 & \texttt{gzip\_then\_base64} & 2 & \texttt{shm\_segment} & 1 \\
\texttt{at\_spool\_job} & 3 & \texttt{hardlink} & 6 & \texttt{signal\_handler} & 1 \\
\texttt{atbash\_cipher} & 1 & \texttt{head\_tail} & 5 & \texttt{sort\_uniq} & 2 \\
\texttt{auth\_log} & 1 & \texttt{hidden\_dotfile} & 9 & \texttt{sparse\_file} & 1 \\
\texttt{awk\_begin\_end} & 5 & \texttt{hmac\_signed\_token} & 4 & \texttt{split\_across\_files} & 6 \\
\texttt{awk\_field} & 15 & \texttt{htpasswd\_hash} & 2 & \texttt{split\_rejoin} & 2 \\
\texttt{background\_process} & 2 & \texttt{iconv\_charset} & 2 & \texttt{sqlite\_blob} & 4 \\
\texttt{base32\_encode} & 1 & \texttt{ini\_config} & 6 & \texttt{sqlite\_query} & 4 \\
\texttt{base64} & 5 & \texttt{iptables\_rules} & 2 & \texttt{ss\_connections} & 2 \\
\texttt{base64\_then\_rot13} & 1 & \texttt{journalctl} & 2 & \texttt{ssh\_keys} & 6 \\
\texttt{base85\_ascii85} & 1 & \texttt{jpeg\_exif\_comment} & 2 & \texttt{sshd\_auth\_keys\_options} & 2 \\
\texttt{bash\_alias} & 3 & \texttt{json\_jq} & 3 & \texttt{sshd\_config\_match} & 2 \\
\texttt{bash\_function} & 2 & \texttt{jwt\_hs256\_token} & 1 & \texttt{stat\_metadata} & 2 \\
\texttt{bash\_history} & 2 & \texttt{ldso\_conf\_d} & 1 & \texttt{sticky\_setuid} & 3 \\
\texttt{bash\_profile} & 8 & \texttt{lsof\_files} & 1 & \texttt{strace\_trace} & 2 \\
\texttt{bc\_math} & 1 & \texttt{mail\_spool} & 1 & \texttt{strings\_binary} & 1 \\
\texttt{bzip2\_file} & 1 & \texttt{md5sum\_verify} & 2 & \texttt{sudoers\_comment} & 2 \\
\texttt{caesar\_shift\_n} & 1 & \texttt{mktemp\_tmpfile} & 2 & \texttt{symlink\_chain} & 5 \\
\texttt{cert\_subject} & 1 & \texttt{mtime\_pre\_epoch} & 3 & \texttt{syslog} & 6 \\
\texttt{chattr\_immutable} & 2 & \texttt{named\_pipe} & 3 & \texttt{systemd\_drop\_in} & 1 \\
\texttt{column\_format} & 3 & \texttt{netcat\_listen} & 5 & \texttt{systemd\_env\_file} & 3 \\
\texttt{comm\_compare} & 3 & \texttt{nl\_numbering} & 6 & \texttt{systemd\_unit\_file} & 2 \\
\texttt{cpio\_archive} & 2 & \texttt{nm\_symbol\_name} & 2 & \texttt{tac\_reverse} & 1 \\
\texttt{cron\_d\_file} & 2 & \texttt{nsswitch\_conf\_source} & 1 & \texttt{tar\_nested} & 5 \\
\texttt{cron\_job} & 3 & \texttt{nsswitch\_lookup} & 6 & \texttt{tee\_duplicate} & 2 \\
\texttt{crontab\_user} & 9 & \texttt{od\_octal} & 2 & \texttt{timeout\_race} & 3 \\
\texttt{csv\_awk} & 7 & \texttt{openssl\_dgst\_salted} & 2 & \texttt{tr\_cipher} & 3 \\
\texttt{cut\_field} & 7 & \texttt{openssl\_enc} & 1 & \texttt{udev\_rule} & 1 \\
\texttt{dd\_extract} & 5 & \texttt{pam\_config} & 5 & \texttt{unicode\_homoglyph} & 2 \\
\texttt{df\_space} & 1 & \texttt{passwd\_shadow} & 2 & \texttt{unix\_socket} & 1 \\
\texttt{diff\_patch} & 3 & \texttt{paste\_merge} & 5 & \texttt{unusual\_location} & 3 \\
\texttt{dmesg\_kernel} & 1 & \texttt{pcre\_regex} & 3 & \texttt{user\_creation} & 6 \\
\texttt{dns\_records} & 2 & \texttt{pdf\_metadata\_info} & 1 & \texttt{uuencode} & 1 \\
\texttt{du\_size} & 2 & \texttt{php\_serialize} & 3 & \texttt{vigenere\_cipher} & 1 \\
\texttt{elf\_gnu\_build\_id} & 3 & \texttt{php\_session\_file} & 1 & \texttt{wc\_counting} & 3 \\
\texttt{elf\_note\_section} & 4 & \texttt{printf\_escape} & 1 & \texttt{whiteout\_overlay} & 1 \\
\texttt{env\_var} & 2 & \texttt{proc\_cmdline} & 1 & \texttt{whitespace\_steg} & 3 \\
\texttt{etc\_environment\_var} & 3 & \texttt{proc\_fd} & 4 & \texttt{x509\_custom\_oid} & 1 \\
\texttt{expand\_tabs} & 2 & \texttt{proc\_maps} & 3 & \texttt{xargs\_exec} & 6 \\
\texttt{file\_capabilities} & 2 & \texttt{profile\_d\_drop\_in} & 2 & \texttt{xattr} & 4 \\
\texttt{filename\_control\_chars} & 4 & \texttt{punycode\_idn} & 2 & \texttt{xml\_grep} & 8 \\
\texttt{filename\_leading\_dash} & 1 & \texttt{quoted\_printable} & 1 & \texttt{xor\_simple} & 3 \\
\texttt{find\_inode} & 1 & \texttt{rev\_reverse} & 1 & \texttt{xxd\_hex} & 7 \\
\texttt{find\_newer} & 6 & \texttt{rot13} & 6 & \texttt{xz\_file} & 1 \\
\texttt{find\_perm} & 7 & \texttt{rot47\_ascii} & 1 & \texttt{yaml\_anchors} & 1 \\
\texttt{find\_size} & 1 & \texttt{ruby\_yaml\_obj} & 1 & \texttt{yes\_timeout} & 1 \\
\texttt{find\_user} & 10 & \texttt{script\_replay} & 3 & \texttt{zip\_password} & 1 \\
\texttt{flac\_vorbis\_comment} & 2 & \texttt{sed\_hold\_space} & 1 & \texttt{zstd\_compress} & 1 \\
\texttt{flock\_lockfile} & 3 & \texttt{sed\_transform} & 1 &  &  \\
\bottomrule
\end{tabular}
\end{table}

\end{document}